\documentclass[10pt,aps,prb,twocolumn,preprintnumbers,amsmath,amssymb,floatfix,showpacs,citeautoscript,longbibliography]{revtex4-1}
\usepackage[latin1]{inputenc}
\usepackage[T1]{fontenc}
\usepackage[english]{babel}
\usepackage[pdftex]{graphicx}
\usepackage{amsmath}
\usepackage{times}
\usepackage[usenames,dvipsnames,svgnames,table]{xcolor}
\usepackage[colorlinks,plainpages=false,linkcolor=blue,urlcolor=blue,citecolor=blue,pdfpagemode=UseNone]{hyperref}

\renewcommand{\AA}{\text{\r{A}}}

\newcommand\Vek[1]{\vec{#1}}

\begin{document}

\title
{
\boldmath
Inducing $n$- and $p$-type thermoelectricity in oxide superlattices by strain tuning of orbital-selective transport resonances
}

\author{Benjamin Geisler}
\email{benjamin.geisler@uni-due.de}
\affiliation{Department of Physics and Center for Nanointegration (CENIDE), Universit\"at Duisburg-Essen, Lotharstr.~1, 47057 Duisburg, Germany}
\author{Rossitza Pentcheva}
\email{rossitza.pentcheva@uni-due.de}
\affiliation{Department of Physics and Center for Nanointegration (CENIDE), Universit\"at Duisburg-Essen, Lotharstr.~1, 47057 Duisburg, Germany}

\date{\today}

\begin{abstract}
By combining first-principles simulations including an on-site Coulomb repulsion term and Boltzmann theory,
we demonstrate how the interplay of quantum confinement and epitaxial strain allows
to selectively design $n$- and $p$-type thermoelectric response
in (LaNiO$_3$)$_3$/(LaAlO$_3$)$_1(001)$ superlattices.
In particular, varying strain from $-4.9$ to $+2.9~\%$ tunes the 
Ni orbital polarization at the interfaces from $-6$ to $+3~\%$.
This is caused by an electron redistribution among Ni $3d_{x^2-y^2}$- and $3d_{z^2}$-derived quantum well states
which respond differently to strain.
Owing to this charge transfer, the position of emerging cross-plane transport resonances can be tuned relative to the Fermi energy.
Already for moderate values of $1.5$ and $2.8~\%$ compressive strain,
the cross-plane Seebeck coefficient reaches $\sim -60$ and $+100$~$\mu$V/K around room temperature, respectively.
This provides a novel mechanism to tailor thermoelectric materials.
Finally, we explore the robustness of the proposed concept with respect to oxygen vacancy formation.
\end{abstract}

\maketitle

\section{Introduction}

Transition metal oxides are attractive for thermoelectric applications
owing to their chemical and thermal stability and environmental friendliness
as well as to the prominent role of electronic correlations~\cite{HeLiuFunahashi:11, HebertMaignan:10}.
Considerable experimental and computational~\cite{Gorai:17} research aims at finding oxide thermoelectrics
with improved performance, mostly among bulk materials
by doping~\cite{Xing:16,Garrity:16,LamontagneGrunerPentcheva:16,Okuda:01} or strain~\cite{Gruner:15}.
An alternative strategy is to exploit heterostructuring and dimensional confinement~\cite{SizeEffectTE:16, HicksDresselhaus:93, Filippetti:12, Delugas:13, Pallecchi:15, GhosezSTO:16, Geisler-LNOSTO:17, GeislerPentcheva-LNOLAO:18}.
This is promoted by the advancement of growth techniques
that allow to design transition metal oxide superlattices (SLs) with atomic precision~\cite{Ohtomo:02, Mannhart:10, Freeland:11, Boris:11, Benckiser:11, ZubkoGhosezTriscone:11, RENickelateReview:16, OxideRoadmap:16}.

One system that has been in the focus of intensive research comprises
the correlated metal LaNiO$_3$ (LNO)
and the wide-gap band insulator LaAlO$_3$ (LAO).
Following the initial proposal of Chaloupka and Khaliullin~\cite{ChaloupkaKhaliullin:08},
significant effort was concentrated on controlling the degree of Ni~$e_g$ orbital polarization
which is absent in bulk LNO.
For instance, Wu \textit{et al.}\ reported an orbital polarization of $-3$ to $+8~\%$
in (LNO)$_4$/(LAO)$_4(001)$ SLs on YAlO$_3$, LaSrAlO$_4$, and SrTiO$_3$ (STO)~\cite{WuBenckiser:13}.
Simulations for ultrathin (LNO)$_1$/(LAO)$_1(001)$ SLs
revealed a similar variation of the orbital polarization between $-4$ and $+2~\%$ for LAO and STO substrates~\cite{ABR:11}.
Variation of the spacer material~\cite{Han-ChemCtrl:10, WuBenckiser:13}
or the crystallographic orientation~\cite{DoennigPickettPentcheva-111-PRB:14}
allows for even higher values.

Several studies have shown that a single or a double LNO layer confined in LAO along the $[001]$ direction
undergoes a metal-to-insulator transition (MIT) for tensile strain~\cite{ABR:11, Freeland:11, LiuChakhalian:11}
and exhibits magnetic order~\cite{Frano:13, Boris:11, Puggioni:12, LuBenckiser:16}.
The origin of this MIT is, however, not related to orbital polarization, as initially proposed,
but to a disproportionation into two inequivalent Ni sites~\cite{ABR:11}.
With increasing LNO thickness,
e.g., in (LNO)$_n$/(LAO)$_3(001)$ SLs on STO ($n=3,5,10$)~\cite{ABR:11, LiuChakhalian:11, GeislerPentcheva-LNOLAO:18}
or (LNO)$_4$/(LAO)$_4(001)$ SLs on STO~\cite{Benckiser:11, LNOLAO-4-4-ParkMillisMarianetti:16},
the metallic behavior of the nickelate is restored.

While the confinement- and strain-induced opening of a small band gap in (LNO)$_1$/(LAO)$_1(001)$ SLs enhances strongly the thermoelectric response,
the thermoelectric performance of (LNO)$_3$/(LAO)$_3(001)$ SLs was found to be impeded by
(i)~the two-dimensional metallic nature of the LNO region and
(ii)~the too thick insulating LAO spacer layer that prohibits vertical transport~\cite{GeislerPentcheva-LNOLAO:18}.
On the other hand, efficient thermoelectric energy conversion requires both $n$- and $p$-type materials
that are structurally and electronically compatible.
To this end, the selective control of interface layer stacking was recently proposed as a strategy
to achieve $n$- and $p$-type response in polar (LNO)$_3$/(STO)$_3(001)$ SLs~\cite{Geisler-LNOSTO:17}.
Designing the thermoelectric response necessitates
quantum control over the spectral transmission function asymmetry around the Fermi energy~\cite{SI86, Geisler-Heusler:15}.
However, the distinct use of transport resonances,
which are attractive for thermoelectric applications~\cite{MahanSofo:96},
has been limited to model studies~\cite{Linke-TE:10, JordanSothmann:13} and materials realizations are lacking so far. 

Here we explore the thermoelectric properties of (LNO)$_3$/(LAO)$_1(001)$ SLs
from first principles.
The single LAO spacer layer
induces the formation of distinct Ni-derived quantum well (QW) states, 
while at the same time permitting electronic transport.
This leads to the emergence of sharp cross-plane transport resonances associated with the $3d_{z^2}$-derived QW states.
Moreover, by using strain the orbital polarization and thus the relative position and occupation of $3d_{x^2-y^2}$- and $3d_{z^2}$-derived QW states
can be tuned.
Interestingly, this mechanism allows to precisely shift the transport resonances
relative to the Fermi energy.
Thereby, considerable $n$- and $p$-type thermoelectric response can be obtained
in one and the same materials combination
by moderately varying the control parameter epitaxial strain
between $-1.5$ and $-2.8~\%$.
We thus exemplify how to control and optimize the thermoelectric properties of oxide heterostructures 
by varying layer thickness, confinement, and epitaxial strain.
Furthermore, we show that the mechanism is robust with respect to oxygen vacancy formation.
This opens a new area of application for the design of orbital polarization,
which is intensively pursued in artificial transition metal oxides.

\begin{figure}[t]
	\centering
	\includegraphics{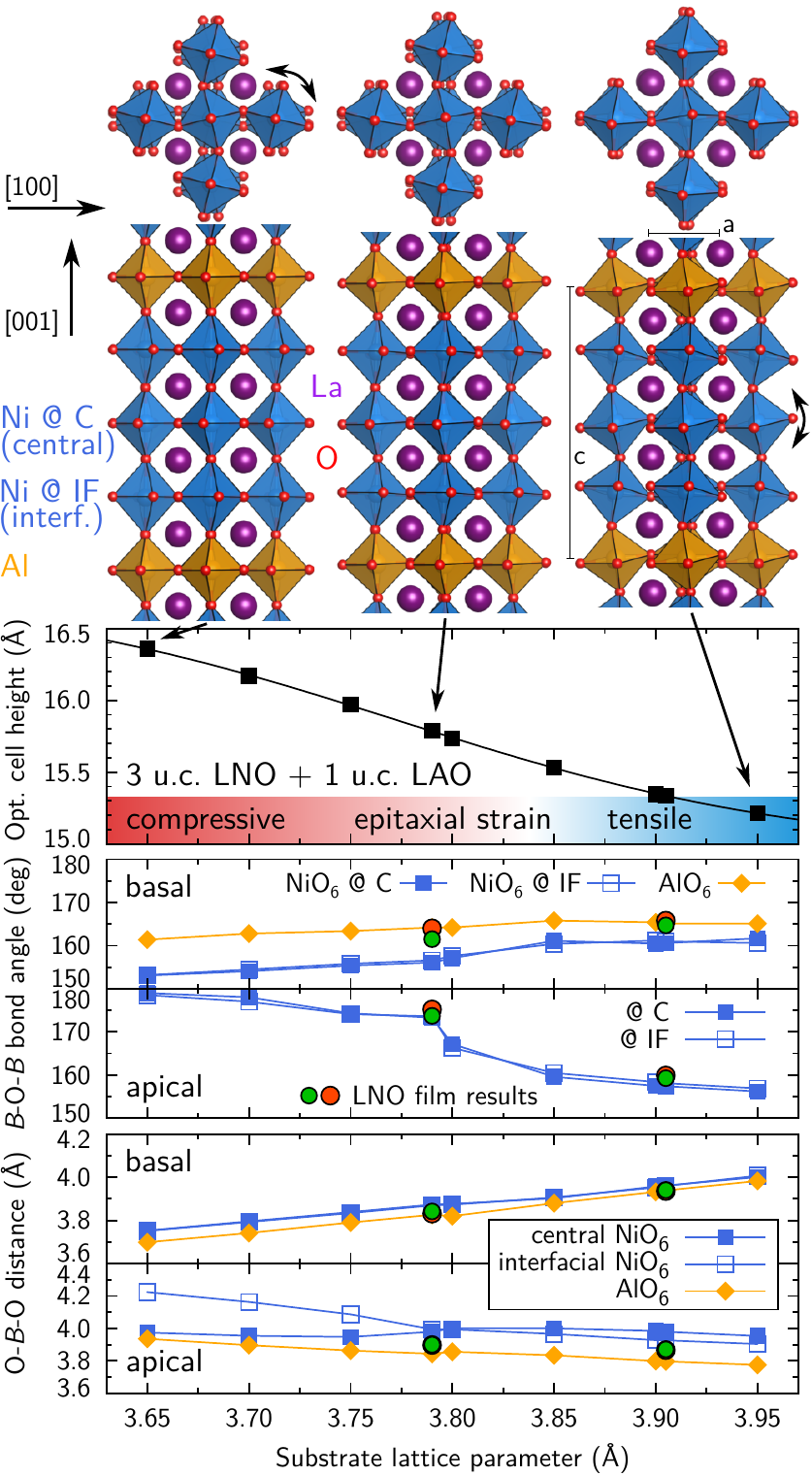}
	\caption{Top and side views of optimized (LNO)$_3$/(LAO)$_1(001)$ SLs for $a=3.65$, $3.79$, and $3.95~\AA$, ranging from compressive (left) to tensile (right) epitaxial strain. Below, several optimized $c(a)$ values are shown, together with a fitted curve. Moreover, $B$-O-$B$ bond angles and O-$B$-O distances are provided as functions of the substrate lattice parameter~$a$. Green (red) circles mark DFT (XRD) results for bulklike LNO films on LAO and STO.~\cite{May:10} Pseudocubic lattice parameters of LAO, LNO, and STO are $3.79$, $3.838$, and $3.905~\AA$, respectively.}
	\label{fig:StructuralProperties}
\end{figure}

\section{Methods}

We performed first-principles calculations in the framework
of spin-polarized density functional theory~\cite{KoSh65} (DFT)
as implemented in Quantum Espresso~\cite{PWSCF}.
The generalized gradient approximation was used for the exchange-correlation functional  
as parametrized by Perdew, Burke, and Ernzerhof~\cite{PeBu96}.
Static correlation effects were considered within DFT$+U$~\cite{Anisimov:93}
using $U=4$ and $J=0.7$~eV for Ni~$3d$,
in line with previous work by us and others~\cite{May:10, ABR:11, KimHan:15, Geisler-LNOSTO:17, WrobelGeisler:18, GeislerPentcheva-LNOLAO:18}.
In order to take octahedral tilts fully into account,
$40$-atom $\sqrt{2}a \times \sqrt{2}a \times c$ supercells were used to 
model the (LNO)$_3$/(LAO)$_1(001)$ SLs.
Epitaxial strain is considered
by varying the in-plane lattice parameter~$a$ from $3.65$ to $3.95~\AA$.
The out-of-plane lattice parameter $c(a)$ was optimized in each case (Fig.~\ref{fig:StructuralProperties}),
as were the atomic positions.
All structures exhibit an antiferrodistortive $a^-a^-c^-$ octahedral rotation pattern
and ferromagnetic order
(see Supplemental Material~\footnote{See Supplemental Material at [xxx] for additional structural data as well as further details concerning the electronic structure and thermoelectric response of (LaNiO$_3$)$_3$/(LaAlO$_3$)$_1(001)$ superlattices.}).
The Brillouin zone was sampled by an $8 \times 8 \times 4$ $\Vek{k}$-point grid~\cite{MoPa76} and $5$~mRy smearing~\cite{MePa89}.
The BoltzTraP code~\cite{BoltzTraP:06} was used to obtain converged energy- and spin-resolved transmission functions ${\cal T}_\sigma(E)$ in Boltzmann theory
(using a denser $64 \times 64 \times 16$ $\Vek{k}$ point grid)
and thermoelectric quantities
according to Sivan and Imry~\cite{SI86},
an approach we have employed also in previous studies~\cite{GeislerPentcheva-LNOLAO:18, Geisler-LNOSTO:17, Geisler-Heusler:15, GeislerPopescu:14, ComtesseGeisler:14};
it is described in the Supplemental Material.

\begin{figure}[t]
	\centering
	\includegraphics{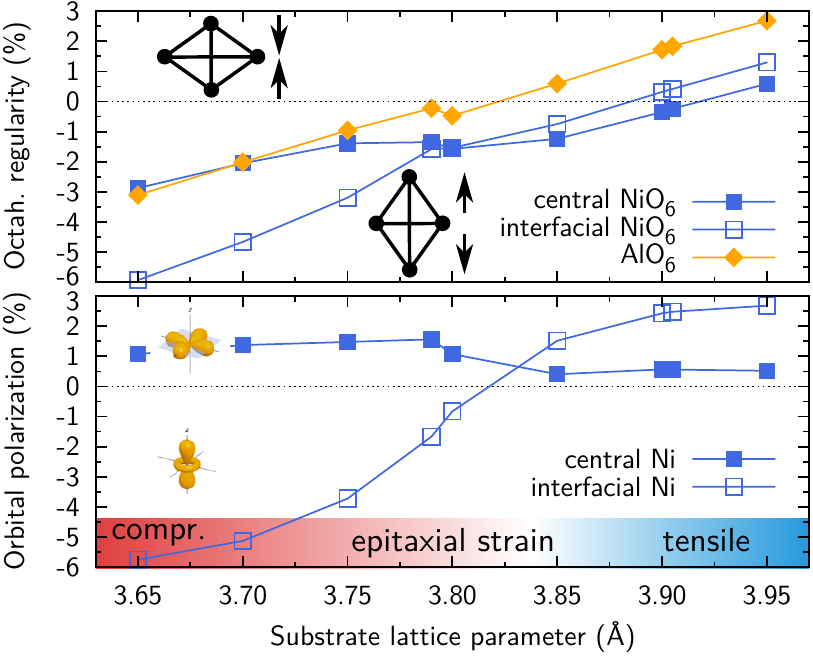}
	\caption{Regularity~$R$ of the NiO$_6$ and AlO$_6$ octahedra (upper panel) and Ni $3d$ $e_g$ orbital polarization~$P$ (lower panel) in (LNO)$_3$/(LAO)$_1(001)$ SLs as functions of~$a$, ranging from compressive (left) to tensile (right) epitaxial strain. Positive (negative) values of $R$ denote a compression (elongation) in the $[001]$ direction, while positive (negative) values of $P$ imply a preferential occupation of the in-plane $3d_{x^2-y^2}$ (out-of-plane $3d_{z^2}$) orbital.}
	\label{fig:OrbitalProperties}
\end{figure}

\begin{figure*}[t]
	\centering
	\includegraphics{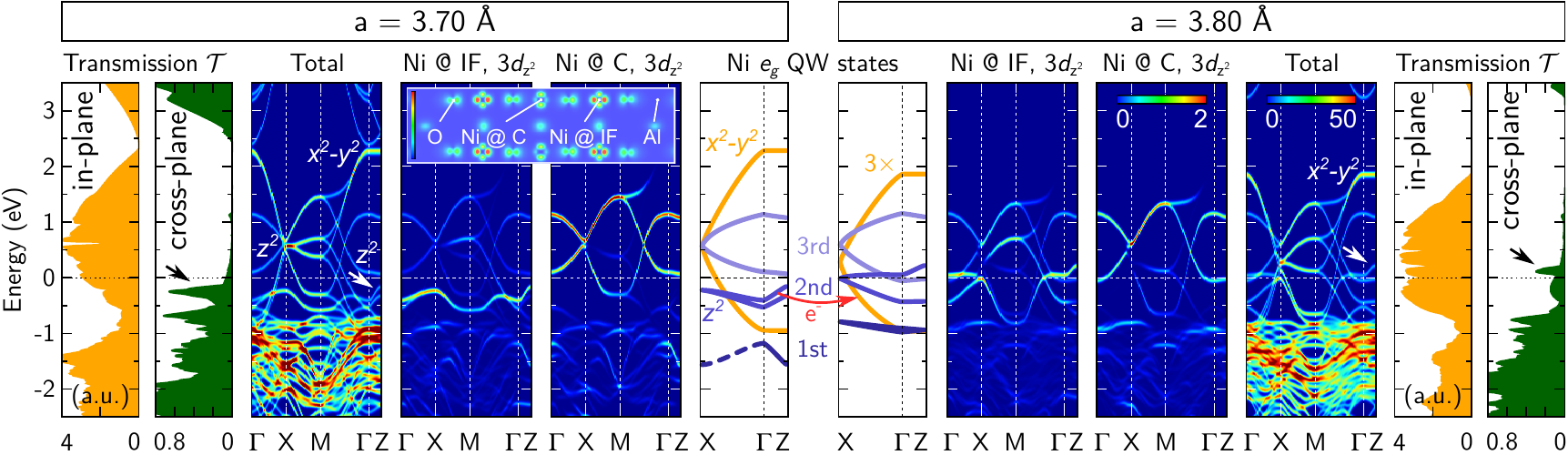}
	\caption{Majority spin band structures (total and projections on $3d_{z^2}$ orbitals at different Ni sites) and related electronic transmission~${\cal T}_\uparrow(E)$ of (LNO)$_3$/(LAO)$_1(001)$ SLs for two selected substrate lattice parameters, $a=3.70$ and $3.80~\AA$. Note the cross-plane transmission peak (black arrows) that stems from Ni $3d_{z^2}$-derived bands along $\Gamma$-$Z$ (i.e., the 2nd QW state; white arrows) and crosses $E_\text{F}$ (zero energy) with increasing epitaxial strain. Analysis of the Ni~$e_g$ QW states reveals that the 1st $3d_{z^2}$-derived QW state exhibits strong contributions from the central nickelate layer, but is also dispersed below the valence band maximum of LAO (which is located $\sim 0.6$~eV below $E_\text{F}$, see Supplemental Material). The 2nd QW state is clearly localized in the interfacial nickelate layers. The 3rd QW state shows contributions from all nickelate layers, but strongest from the central one. The inset displays the local density of states~\cite{Geisler-TMs:15, Geisler:13, Geisler:12} integrated from $-0.5$ to $-0.3$~eV below $E_\text{F}$, visualizing predominantly the 2nd QW state. The central panels emphasize schematically the evolution of the different $3d_{x^2-y^2}$- (orange, quasi threefold degenerate) and $3d_{z^2}$-derived (blue) QW states with epitaxial strain. -- The band structures correspond to $\smash{\Vek{k}}$-resolved densities of states, in which each electronic state is represented by a broadened delta distribution of weight one (total) or a weight equal to the projection of the respective wave function on $3d_{z^2}$ orbitals at different Ni sites (projected). The color scales are in units of $1/$eV.}
	\label{fig:BandsTransmissionQW}
\end{figure*}

\section{Interplay of structure and orbital polarization}

Figure~\ref{fig:StructuralProperties} shows
the optimized (LNO)$_3$/(LAO)$_1(001)$ SL structures
and cell heights~$c(a)$
for several in-plane lattice parameters~$a$,
modeling the growth on a variety of substrates
imposing compressive strain
[such as ScAlO$_3$ ($3.60~\AA$), YAlO$_3$ ($3.716~\AA$), or LAO ($3.79~\AA$)]
or tensile strain
[such as STO ($3.905~\AA$), DyScO$_3$ ($3.94~\AA$), or GdScO$_3$ ($3.976~\AA$)]~\cite{LandoltBoernstein-Perovskites, Ross:98}.
The relaxed $c(a)$ parameters for discrete $a$ values were fitted to
$c(a) = c_0 \tanh \left\lbrace \tilde{a} (a-a_\text{i}) \right\rbrace + c_1$,
rendering 
$c_0 = -0.934~\AA$,
$c_1 = 15.86~\AA$,
$\tilde{a} = 4.818/\AA$, and
$a_\text{i} = 3.773~\AA$ (inflection point).
With the pseudocubic lattice parameter of LNO, $a_\text{LNO} = 3.838~\AA$~\cite{May:10},
this corresponds to degrees of epitaxial strain $\epsilon = a/a_\text{LNO}-1$
ranging from $-4.9$ to $+2.9~\%$.
The strong octahedral rotations around the $c$~axis
present for compressive strain
are continuously reduced
while shifting to tensile strain.
This is reflected in part in the basal $B$-O-$B$ bond angles
that increase from $153^\circ$ to $161^\circ$ (Ni)
and from $161^\circ$ to $165^\circ$ (Al;
$180^\circ$ implying the absence of any rotation).
In contrast, while
almost no octahedral rotations around the $a$~axes are present for strong compressive strain,
they appear at $a \approx 3.80~\AA$,
signaled by a sudden decrease of the apical $B$-O-$B$ bond angles
(stronger tilts) from $\sim 175^\circ$ to $165^\circ$.
This onset is located close to the inflection point~$a_\text{i}$
of the $c(a)$ curve;
the relaxation pattern around this point has been carefully checked to avoid local minima.
The basal O-Ni-O and O-Al-O distances increase from $3.7$ to $4.0~\AA$,
whereas the apical O-Al-O and interfacial O-Ni-O distances decrease from $3.95$ to $3.78~\AA$ and from $4.2$ to $3.9~\AA$, respectively.
The central apical O-Ni-O distance is almost constant ($\sim 4~\AA$).
Comparison of the nickelate region
to DFT and x-ray diffraction (XRD) results for $100$-$200~\AA$ thick LNO films grown on LAO and STO substrates~\cite{May:10}
reveals major similarities,
e.g., the strong variation of the apical bond angles with~$a$ (Fig.~\ref{fig:StructuralProperties}).
Similar to the (LNO)$_3$/(LAO)$_3(001)$ SLs
and at variance with the (LNO)$_1$/(LAO)$_1(001)$ SLs~\cite{GeislerPentcheva-LNOLAO:18},
the present system shows no tendency towards an in-plane Ni-site disproportionation
and remains metallic for all considered substrate lattice constants.

We quantify the octahedral regularity~$R$ and the Ni~$3d$ $e_g$ orbital polarization~$P$
(from O-$B$-O distances~$d$ and Ni orbital occupations~$n$)
shown in Fig.~\ref{fig:OrbitalProperties}
by the expressions
\begin{equation*}
R = \frac{d_\text{O-$B$-O}^\text{ basal} - d_\text{O-$B$-O}^\text{ apical}}{d_\text{O-$B$-O}^\text{ basal} + d_\text{O-$B$-O}^\text{ apical}}
\ \
\text{and}
\ \
P = \frac{ n(3d_{x^2-y^2}) - n(3d_{z^2}) }{ n(3d_{x^2-y^2}) + n(3d_{z^2}) }
\text{.}
\end{equation*}
Positive (negative) values of $R$ denote a compression (elongation) in the $[001]$ direction,
while positive (negative) values of $P$
imply a preferential occupation of the in-plane $3d_{x^2-y^2}$ (out-of-plane $3d_{z^2}$) orbital.
We reduce projection errors arising from the tilts of the octahedra
by applying a simple correction scheme to the 'Cartesian' orbital occupations
(see Supplemental Material).
While $P$ is almost constant in the central LNO layer ($\sim +1~\%$),
it changes from $-6~\%$ (compressive strain) to $\sim +3~\%$ (tensile strain)
in the interfacial LNO layers.
Likewise, $R$ of the interfacial NiO$_6$ octahedra shows the strongest strain dependence and varies between $-6$ and $+1.25~\%$.
Experiments on (LNO)$_4$/(LAO)$_4(001)$ SLs found a similar variation of $P$ with strain~\cite{WuBenckiser:13}.
We thus conclude from Fig.~\ref{fig:OrbitalProperties}
that the interfacial Ni orbital polarization can be designed considerably by strain,
displaying the strongest variation around $a=3.80~\AA$.

The evolution of the electronic structure with epitaxial strain
is shown exemplarily for $a=3.70$ and $3.80~\AA$ in Fig.~\ref{fig:BandsTransmissionQW}.
As previously observed in (LNO)$_3$/(STO)$_3(001)$ and (LNO)$_3$/(LAO)$_3(001)$ SLs~\cite{Geisler-LNOSTO:17, GeislerPentcheva-LNOLAO:18},
distinct Ni $3d_{x^2-y^2}$- and $3d_{z^2}$-derived QW states form
within the band gap of the spacer material due to confinement. 
While the $3d_{x^2-y^2}$-derived QW states are restricted to single LNO layers and show almost no influence of the LAO spacer layer,
the $3d_{z^2}$-derived QW states extend over the entire QW and split up strongly.
Consequently, the two sets of states have a distinct response to a variation of the substrate lattice parameter:
With increasing strain, the $3d_{x^2-y^2}$- \mbox{($3d_{z^2}$-)} derived QW states are shifted to lower (higher) energies.
This impacts the band occupations:
Figure~\ref{fig:BandsTransmissionQW} shows that
the strong increase of $P$ in the interfacial LNO layers with strain
(Fig.~\ref{fig:OrbitalProperties})
is caused by a
charge transfer
from the localized 2nd $3d_{z^2}$-derived QW state 
to the set of dispersive $3d_{x^2-y^2}$-derived QW states.
(More information can be found in the Supplemental Material.)

\begin{figure}[t]
	\centering
	\includegraphics{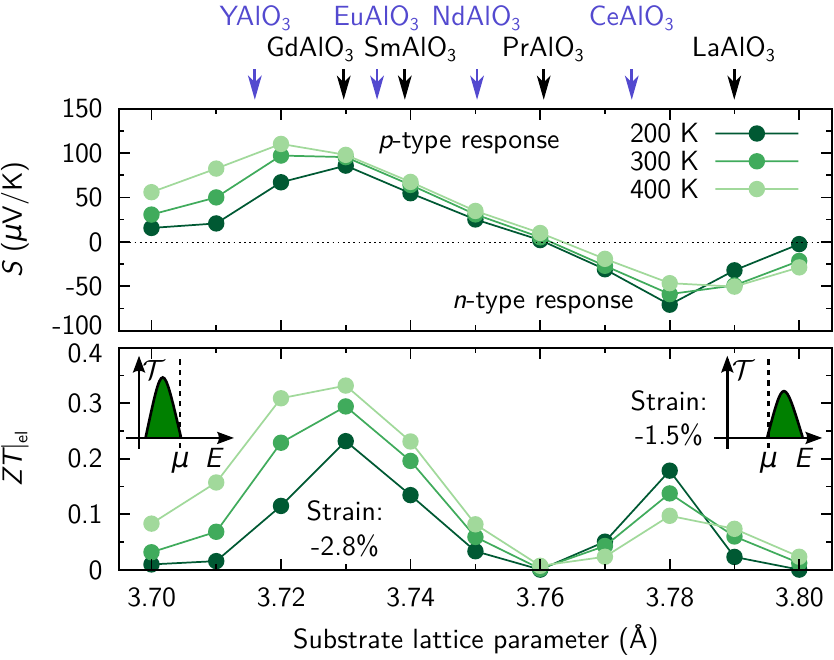}
	\caption{Thermoelectric properties of (LNO)$_3$/(LAO)$_1(001)$ SLs at $\mu = E_\text{F}$ and three different temperatures for cross-plane transport as functions of~$a$. Epitaxial strain as imposed by exemplary (rare-earth) aluminate substrates is marked (calculated from Ref.~\onlinecite{LandoltBoernstein-Perovskites}). The insets depict the evolution of the transmission peak with strain (cf.~Fig.~\ref{fig:BandsTransmissionQW}).}
	\label{fig:ThermoelectricProperties}
\end{figure}

\section{Thermoelectric response exploiting transport resonances}

The single LAO spacer layer allows for a significant cross-plane electronic transmission (Fig.~\ref{fig:BandsTransmissionQW}).
In particular,
resonances emerge, which originate from the $3d_{z^2}$-derived QW states
and whose position
can be shifted considerably
with respect to the Fermi energy~$E_\text{F}$
by epitaxial strain.
We emphasize that
this is only possible due to the opposite response of $3d_{x^2-y^2}$- and $3d_{z^2}$-derived QW states to strain,
which causes the strong change in orbital polarization (Fig.~\ref{fig:OrbitalProperties})
by a redistribution of electrons (Fig.~\ref{fig:BandsTransmissionQW});
otherwise, $E_\text{F}$ would simply shift with the resonances.
Particularly, the cross-plane transmission peak close to $E_\text{F}$,
which stems from Ni $3d_{z^2}$-derived bands along $\Gamma$-$Z$
(i.e., the 2nd QW state),
shifts from $-0.24$ to $+0.12$~eV
as $a$ increases from $3.70$ to $3.80~\AA$.
Since the Seebeck coefficient~$S$ is highly sensitive to the asymmetry of the transmission
around~$\mu \sim E_\text{F}$,
this resonance can be exploited for a targeted design of the thermoelectric response:
For $300$~K, we find that $S$ varies between $+100$ ($a = 3.73~\AA$, similar to YAlO$_3$, GdAlO$_3$ substrates)
and $-60$~$\mu$V/K ($a = 3.78~\AA$, similar to CeAlO$_3$, LAO substrates);
at the same time, moderate values are obtained for the electronic figure of merit $ZT|_\text{el} = \sigma S^2 T / \kappa_\text{el}$,
attaining $0.30$ and $0.14$ (Fig.~\ref{fig:ThermoelectricProperties}, Table~\ref{tab:TE-Summary}).
Thus, epitaxial strain acts as a control parameter
that allows to induce $n$- ($S<0$) and $p$-type ($S>0$) thermoelectric response in one and the same materials combination.

In the constant (i.e., energy-independent) relaxation time approximation,
$S$ and $ZT|_\text{el}$ do not depend on~$\tau$.
Since in this work the focus lies on a narrow energy window (around the transmission peak),
we consider this approximation to be well-justified.
Around room temperature, electron-phonon scattering contributes substantially to~$\tau$~\cite{GeislerPentcheva-LNOLAO:18}
and might also exhibit a strain dependence.
While $ZT|_\text{el}$ renders the contribution of the electronic system,
the phonon heat conductivity~$\kappa_\text{ph}$ needs to be considered to obtain the total $ZT$.
We note, however, that for the SLs studied here the respective bulk values
($\kappa_\text{ph} \sim 0.035$~W/K\,cm for bulk LNO~\cite{LNO-LCO-Thermo:14} and $\sim 0.1$~W/K\,cm for bulk LAO~\cite{Schnelle:01} at $300$~K)
may not be the relevant quantities, since the presence of interfaces is expected to have a favorable effect on~$\kappa_\text{ph}$ due to scattering of phonons.

The thermoelectric performance is higher for the $p$-type case than for the $n$-type case,
since the intensity of the transmission peak increases from $a=3.80$ to $3.70~\AA$ (Fig.~\ref{fig:BandsTransmissionQW})
owing to an enhanced dispersion along $\Gamma$-$Z$.
The $n$- and $p$-type response exhibit opposite dependence on temperature: while the former decreases, the latter increases.
The system shows strong anisotropy,
$S$ being negligible in-plane (not shown here) due to the two-dimensional metallic character (Fig.~\ref{fig:BandsTransmissionQW}).

Comparison with literature (Table~\ref{tab:TE-Summary}) reveals that
the thermoelectric properties are strongly improved
in the (LNO)$_3$/(LAO)$_1(001)$ SL 
with respect to bulk LNO and (LNO)$_3$/(LAO)$_3(001)$ SLs~\cite{GeislerPentcheva-LNOLAO:18}
owing to the reduction to a single LAO spacer layer.
The system attains $S$ and $ZT|_\text{el}$ values that are
similar in magnitude to those of metallic (LNO)$_1$/(LAO)$_1(001)$ SLs at $a_\text{LAO}$
or (LNO)$_3$/(STO)$_3(001)$ SLs at $a_\text{STO}$~\cite{GeislerPentcheva-LNOLAO:18, Geisler-LNOSTO:17}.
Only (LNO)$_1$/(LAO)$_1(001)$ SLs at $a_\text{STO}$,
which undergo a MIT for tensile strain~\cite{ABR:11},
show better values that are comparable to the best oxide thermoelectrics~\cite{GeislerPentcheva-LNOLAO:18}.
Another beneficial feature of the LAO layers is that they
are expected to scatter LNO phonons due to the high mass difference,
thereby reducing the detrimental cross-plane lattice thermal conductivity.

\begin{table}[b]
	\centering
	\caption{\label{tab:TE-Summary}Comparison of the attainable thermoelectric performance of the present system around $300$~K (Fig.~\ref{fig:ThermoelectricProperties}) to related nickelate SLs (DFT$+U$ results) and bulk LNO, i.e., without heterostructuring. (See Supplemental Material for further information.)}
	\begin{ruledtabular}
	\begin{tabular}{lccr}
		$a$ ($\AA$)	& $\epsilon$ (\%)		& $S$ ($\mu$V/K)	& $ZT|_\text{el}$	\\
		\hline
		\multicolumn{4}{l}{(LNO)$_3$/(LAO)$_1(001)$ SL, cross-plane (this work)}	\\
		$3.73$		& $-2.8$				& $+100$				& $0.30$	\\
		$3.78$		& $-1.5$				& $-60$					& $0.14$	\\
		\hline
		\multicolumn{4}{l}{(LNO)$_1$/(LAO)$_1(001)$ SL, cross-plane~\cite{GeislerPentcheva-LNOLAO:18} }	\\
		$3.79 = a_\text{LAO}$				& $-1.3$				& $-80$			& $0.2$	\\
		$3.905 = a_\text{STO}$				& $+1.7$				& $\pm 600$		& $0.8$-$1.0$	\\
		\hline
		\multicolumn{4}{l}{(LNO)$_3$/(LAO)$_3(001)$ SL, cross-plane~\cite{GeislerPentcheva-LNOLAO:18} }	\\
		$3.79 = a_\text{LAO}$				& $-1.3$				& $-5$, $+15$	& $<0.01$	\\
		$3.905 = a_\text{STO}$				& $+1.7$				& $-5$, $+15$	& $<0.01$	\\
		\hline
		\multicolumn{4}{l}{(LNO)$_3$/(STO)$_3(001)$ SL, cross-plane, $p$-type interfaces~\cite{Geisler-LNOSTO:17} }	\\
		$3.905 = a_\text{STO}$				& $+1.7$				& $+135$		& $0.35$	\\
		\hline
		\multicolumn{2}{l}{LNO bulk (DFT$+U$, this work)}				& $-12$				& $0.006$	\\
		\multicolumn{2}{l}{LNO bulk (exp.~\cite{LNO-LCO-Thermo:14}, $\kappa_\text{total}$)}				& $-18$				& $0.009$	\\
		\multicolumn{2}{l}{LNO bulk (exp.~\cite{LNO-LCO-Thermo:14}, $\kappa_\text{Wiedemann-Franz}$)}	& $-18$				& $0.013$	\\
	\end{tabular}
	\end{ruledtabular}
\end{table}

\begin{figure}[t]
	\centering
	\includegraphics{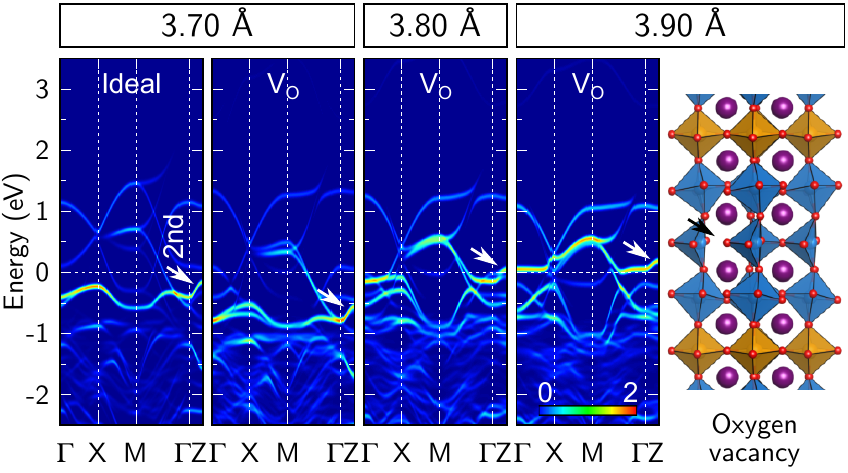}
	\caption{Impact of oxygen vacancies in (LNO)$_3$/(LAO)$_1(001)$ SLs at three different substrate lattice constants. The projected majority spin band structures (interfacial Ni $3d_{z^2}$ orbitals) show that the released electrons are accommodated by the $3d_{x^2-y^2}$-derived QW states, leading to a shift of $E_\text{F}$ (zero energy; cf.~Fig.~\ref{fig:BandsTransmissionQW}), and that the $3d_{z^2}$-derived QW states, which give rise to the transport resonances, are preserved. The white arrows mark the 2nd QW state, which crosses $E_\text{F}$ at $\sim 3.80~\AA$. On the right, a representative optimized SL geometry at $a = 3.90~\AA$ is shown (cf.~Fig.~\ref{fig:StructuralProperties}).}
	\label{fig:OxVacs}
\end{figure}

\section{Impact of oxygen vacancies}

Finally,
we explore the robustness of our proposed mechanism with respect to the formation of oxygen vacancies (Fig.~\ref{fig:OxVacs}).
To this end, we generated an isolated vacancy in our supercell for three substrate lattice constants~$a$,
corresponding to a (high) V$_\text{O}$ concentration of $4.2$~\% of the oxygen sites.
We find that the released electrons are accommodated by the $3d_{x^2-y^2}$-derived QW states,
leading effectively to a shift of $E_\text{F}$ (Fig.~\ref{fig:OxVacs}).
Most importantly, the $3d_{z^2}$-derived QW states, which give rise to the transport resonances, are preserved.
The resonance corresponding to the 2nd QW state crosses $E_\text{F}$ at a substrate lattice parameter of $a \sim 3.80$ instead of $\sim 3.76~\AA$ (cf.~Fig.~\ref{fig:ThermoelectricProperties}).
Hence, under the presence of oxygen vacancies, slightly higher strain is required to achieve a selective design of $n$- and $p$-type thermoelectric response as in the case of an ideal SL.
The vacancy formation energies under oxygen-rich conditions can be derived from DFT total energies,
\begin{equation*}
  E_{\text{V}_\text{O}}^\text{f} = E({\text{SL with V}_\text{O}}) - E(\text{SL, ideal}) + \tfrac{1}{2} E({\text{O}_2}) \ \text{,}
\end{equation*}
and amount to $E_{\text{V}_\text{O}}^\text{f} = 2.25$, $2.15$, and $2.29$~eV for $a = 3.70$, $3.80$, and $3.90~\AA$, respectively (endothermic).
Comparison to the DFT value for bulk LNO of $E_{\text{V}_\text{O}}^\text{f} = 2.8 \pm 0.2$~eV~\cite{LNO-OxVac-Beigi:15}
suggests that our SLs are slightly more prone to oxygen vacancy formation
\footnote{The well-known overbinding of gas-phase O$_2$ molecules in DFT~\cite{LNO-OxVac-Beigi:15}
necessitates a correction of $E({\text{O}_2})$,
which we performed such as to reproduce the experimental O$_2$ binding energy of $5.16$~eV.}.

\section{Summary}

By combining density functional theory calculations with an on-site Coulomb repulsion term
and Boltzmann theory in the constant relaxation time approximation,
we explored the thermoelectric properties of (LaNiO$_3$)$_3$/(LaAlO$_3$)$_1(001)$ superlattices
and the impact of different substrate lattice parameters.
Variation of epitaxial strain between $-4.9$ and  $+2.9~\%$
causes an electron redistribution among Ni $3d_{x^2-y^2}$- and $3d_{z^2}$-derived quantum well states
that results in a considerable change in the orbital polarization ($-6$ to $+3~\%$).
Concomitantly,
sharp cross-plane transport resonances emerging from the $3d_{z^2}$-derived quantum well states,
induced by the confinement due to the single LaAlO$_3$ spacer layer,
can be shifted relative to the Fermi energy.
Already slight variation of the control parameter epitaxial strain
between $-1.5$ and $-2.8~\%$
is sufficient
to selectively induce a considerable $n$- and $p$-type thermoelectric response
($\sim -60$ and $+100$~$\mu$V/K)
in one and the same oxide superlattice.
We showed that oxygen vacancies cause only a small shift of this crossover point.
This constitutes an interesting mechanism and a concrete materials realization in an oxide system of current interest
that calls for further experimental and theoretical exploration.
We expect that the results can be extended to a broader class of oxide heterostructures.

\begin{acknowledgments}

This work was supported by the German Science Foundation (Deutsche Forschungsgemeinschaft, DFG) within the SFB/TRR~80, projects G3 and G8.
Computing time was granted by the Center for Computational Sciences and Simulation of the University of Duisburg-Essen
(DFG grants INST 20876/209-1 FUGG, INST 20876/243-1 FUGG).

\end{acknowledgments}


%

\end{document}